\documentstyle[12pt,epsf]{article}
\begin{document}

\begin{center}
\large{\bf Chiral Symmetry and the Nucleon Spin Structure
Functions\footnote{Invited talk at the XIV International Seminar on High Energy Physics 
Problems ``Relativistic Nuclear Physics and Quantum Chromodynamics'', 
Dubna, 17-22 August, 1998
}}

\vskip 5mm
M.~Wakamatsu$^{\dag}$

\vskip 5mm

{\small

{\it Department of Physics, Faculty of Science,}\\
{\it Osaka University, Toyonaka, Osaka 560, Japan}
\\
$\dag$ {\it
E-mail: wakamatu@miho.rcnp.osaka-u.ac.jp}
}
\end{center}

\vskip 4mm

\begin{center}
\begin{minipage}{150mm}
\centerline{\bf Abstract}
\begin{small}
\ \ \ We carry out a systematic investigation of twist-two spin
dependent structure functions of the nucleon within the framework
of the chiral quark soliton model (CQSM) by paying special
attention to the role of chiral symmetry of QCD. We observe
a substantial difference between the predictions of the
longitudinally polarized distribution functions and the
transversity distribution ones. That the chiral symmetry is
responsible for this difference can most clearly be seen in the
isospin dependence of the corresponding first moments, i.e.
the axial and tensor charges. The CQSM predicts
$g_A^{(0)} / g_A^{(3)} \simeq 0.25$ for the ratio of the
isoscalar to isovector axial charges, and $g_T^{(0)} /
g_T^{(3)} \simeq 0.46$ for the ratio of the isoscalar to
isovector tensor charges, which should be compared with the
prediction of the naive (non-chiral) MIT bag model,
$g_A^{(0)} / g_A^{(3)} = g_T^{(0)} / g_T^{(3)} = 3 / 5$.
\end{small}
\end{minipage}
\end{center}

\vspace{4mm}
\noindent
\begin{large}
{\bf 1. Introduction}
\end{large}
\vspace{3mm}

\ \ \ Undoubtedly, ``Nucleon Spin Crisis'' caused by
the EMC measurement in 1988 is one of the most exciting topics
in the field of hadron physics [1]. The recent renaissance of nucleon
structure function physics is greatly owing to this
epoch-making finding. Here we recall that the physics of nucleon
structure functions has two different aspects. One is a
perturbative aspect. Because of the asymptotic freedom of QCD,
the $Q^2$-evolution of quark distribution functions can be
controlled by the perturbative QCD at least for large enough $Q^2$.
However, the perturbative QCD is totally powerless for predicting
distribution functions themselves. Here we need to solve
nonperturbative QCD in some way. Unfortunately, we have no reliable
analytical method yet. We are then left with two tentative choices.

One is to rely upon lattice QCD, while the other is to use effective
models of QCD. If one takes the first choice, one must first
calculate infinite towers of moments of distribution functions,
since the direct evaluation of distribution functions does not
match this numerical method. Here we take the second choice,
which allows us a direct evaluation of quark distribution
functions. Naturally, there are quite a lot of effective model
of baryons. But let me shortly explain merits of our model,
i.e. the chiral quark soliton model (CQSM) over the others.
The CQSM is an effective model of baryons maximally incorporating
$\chi$SB of QCD vacuum [2,3]. The nucleon in this model is a composite
of 3 valence quarks and infinitely many sea quarks moving in a
slowly rotating M.F. of hedgehog shape. It automatically simulates
cloud of pions surrounding the core of three valence quarks.
Noteworthy here is that our pion fields are not
independent fields of quarks, but they are rather $q \bar{q}$
composites. Since everything is described in terms of effective
quark fields, we need not worry about a double
counting of quark and pion degrees of freedom. This also means
that we have no necessity of convoluting pion structure functions
with pion probability function inside the nucleon [4-6].

\vspace{5mm}
\noindent
\begin{large}
{\bf 2. CQSM and quark distribution functions}
\end{large}
\vspace{3mm}

\ \ \ We start with a definition of quark distribution function given
as a Fourier transform of the nucleon matrix element of bilocal quark
operator [7] :
\begin{eqnarray*}
   q (x) \ \ &=& \ \ \frac{1}{4 \pi} \int^{\infty}_{-\infty} 
   d z^0 \,e^{i x M_N z^0} \\
   &\times& <N (\mbox{\boldmath $P$}=0) \,| \,
   \psi^\dagger (0) \,O \,
   \psi(z) \,| \,N 
   (\mbox{\boldmath $P$}=0)> \,\,
   |_{z^3 = -z^0,\, z_{\perp} = 0} \,\, .
\end{eqnarray*}
Here the constraint $z^3 = - z^0, z_{\perp} = 0$ for
the space time coordinates means that
we need to evaluate quark-quark correlation function with
light-cone separation. In the present talk, let me confine to spin
dependent distribution functions of leading twist 2. There are two
kinds of twist-2 spin dependent distribution functions [8]. One is
the familiar longitudinally polarized distribution functions. The
other is the less familiar chiral-odd distribution functions, which
are sometimes called the transversity distribution functions.
Both distribution functions consist of isoscalar part and
isovector part. They are obtained by inserting the following
operator into the above general expression of quark distribution
functions :
\begin{eqnarray}
   \Delta u(x) \ + \ \Delta d(x) \ \ \ &:& \ \ \ 
   O \ = \ (\,1 + \gamma^0 \gamma^3) \,\gamma_5 \,\, ,\\
   \Delta u(x) \ - \ \Delta d(x) \ \ \ &:& \ \ \ 
   O \ = \ \tau_3 \,
   (\,1 + \gamma^0 \gamma^3) \,\gamma_5 \,\, ,
\end{eqnarray}
for the longitudinally polarized distributions, and
\begin{eqnarray}
   \delta u(x) \ + \ \delta d(x) \ \ \ &:& \ \ \ 
   O \ = \ (\,1 + \gamma^0 \gamma^3) \,
   \gamma_{\perp} \,\gamma_5 \,\, ,\\
   \delta u(x) \ - \ \delta d(x) \ \ \ &:& \ \ \ 
   O \ = \ \tau_3 \,
   (\,1 + \gamma^0 \gamma^3) \,
   \gamma_{\perp} \,\gamma_5 \,\, .
\end{eqnarray}
for the transversity distributions.
The basis of our theoretical analysis is the following path integral
representation of the nucleon matrix element of bilocal quark
operator :
\begin{eqnarray}
   &\,& <N (\mbox{\boldmath $P$}) \,| \,\psi^\dagger (0) \,O \,
   \psi(z) \,| \,N (\mbox{\boldmath $P$})> \\
   &=& \frac{1}{Z} \,\,\int \,\,d^3 x  \,\,d^3 y \,\,
   e^{\,- \,i \mbox{\boldmath $P$} \cdot \mbox{\boldmath $x$}} \,\,
   e^{\,i \,\mbox{\boldmath $P$} \cdot \mbox{\boldmath $y$}} \,\,
   \int {\cal D} \mbox{\boldmath $\pi$} \,\,\int {\cal D} \,\,
   \psi \,\,{\cal D} \,\,\psi^\dagger \\
   \hspace{-6mm} &\times& \!\!\
   J_N (\frac{T}{2}, \mbox{\boldmath $x$}) \,\cdot \,
   \psi^\dagger (0) \,\,
   O \,\,\psi(z) \,\cdot \,J_N^\dagger (-\frac{T}{2}, 
   \mbox{\boldmath $y$})
   \,\,\,\, e^{\,i \int \,d^4 x \,\,{\cal L}_{CQM}} \,\, ,
\end{eqnarray}
with
\begin{equation}
   {\cal L}_{CQM} \ \ = \ \ \bar{\psi} \,(\,i \not\!\partial \ - \ 
   M \,e^{\,i \gamma_5 \mbox{\boldmath $\tau$} \cdot
   \mbox{\boldmath $\pi$} (x) / f_\pi \,} \,) \,\psi \,\, ,
\end{equation}
being the basic lagrangian of the CQSM.
We start with a stationary pion field configuration of hedgehog
form :
\begin{equation}
   \mbox{\boldmath $\pi$} (x) \ = \ f_\pi \,
   \hat{\mbox{\boldmath $r$}} \, F(r) \,\, .
\end{equation}
Next we carry out a path integral over
$\mbox{\boldmath $\pi$} (x)$ in a saddle point approximation by
taking care of two zero-energy modes, i.e. the translational
zero-modes and rotational zero-modes. Under the assumption of
``slow rotation'' as compared with the intrinsic quark motion, the
answer can be obtained in a perturbative series in the collective
rotational velocity $\Omega$. Without going into the detail, here
I just mention that the isoscalar distribution receives only the
$O (\Omega^1)$ contribution, while the isovector distribution
contains both of $O (\Omega^0)$ and $O (\Omega^1)$
contributions [9] :
\begin{eqnarray}
   \Delta u(x) \ + \ \Delta d(x) &\sim& \ \ \ \ 0
   \ \ \ \ \,\,+ \ O(\Omega^1) \,\, ,\\
   \Delta u(x) \ - \ \Delta d(x) &\sim& \ 
   O(\Omega^0) \ + \ O(\Omega^1) \,\, .
\end{eqnarray}
Since the model contains ultraviolet divergences, it needs
a regularization. This can most easily be seen from the effective
meson action obtained from our basic lagrangian :
\begin{eqnarray}
   S_{eff} [U] &=& - \,i \,N_c \,\mbox{Sp} \,\log
   \,[ \,i \not\!\partial - M U^{\gamma_5} ] \\
   &=& \ \frac{4 N_c}{f_{\pi}^2} \,\,I_2 (M) \cdot 
   \frac{1}{2} \,(\partial_{\mu} \mbox{\boldmath $\pi$})^2 
   \ + \ \cdots \,\, ,
\end{eqnarray}
with
\begin{eqnarray}
   \ \ \ I_2 (M) &\equiv& i \,\int \frac{d^4 k}{(2 \pi)^4} \,
   \frac{M^2}{(k^2 - M^2)^2} \,\, .
\end{eqnarray}
The coefficient of pion kinetic term is logarithmically
divergent. To regularize it, here we make use of the Pauli-Villars
subtraction scheme [4]. That is, we subtract from the original action
another action in which the constituent quark mass $M$ is replaced
by a Pauli-Villars mass $M_{PV}$ :
\begin{equation}
   S_{eff}^{reg} \ \equiv \ S_{eff}^M \ - \ 
   {\left(\frac{M}{M_{PV}}\right)}^2 \,
   S_{eff}^{M_{PV}} \,\, .
\end{equation}
This subtraction makes the
coefficient of pion kinetic term finite. If we further require
this kinetic term has a correct normalization, the Pauli-Villars
mass is uniquely fixed from the equation :
\begin{equation}
   \frac{N_c}{4 \,\pi^2} \,M^2 \,\log \,
   {\left(\frac{M}{M_{PV}}\right)}^2 \ = \ 
   f_\pi^2 \,\, .
\end{equation}
Other observables like quark distribution
functions, which contain logarithmic divergence, can similarly be
regularized by using the same subtraction :
\begin{equation}
   <O>^{reg} \ \ \equiv \ \ <O>^M \ - \ 
   {\left(\frac{M}{M_{PV}}\right)}^2 <O>^{M_{PV}} \,\, .
\end{equation}
%

\vspace{5mm}
\noindent
\begin{large}
{\bf 3. Results and Discussion}
\end{large}
\vspace{3mm}

\ \ \ Before showing the results of our numerical calculation, we
briefly mention parameters of the model. The most important parameter
of the CQSM is the dynamical quark mass $M$, which plays the role
of the quark-pion coupling constant. Here we simply use the value
$ M = 375 \,\mbox{MeV}$ established from our previous analyses
of baryon observables.
As already mentioned, the Pauli-Villars mass as a regularization
parameter is uniquely fixed to be $M_{PV} = 562 \,\mbox{MeV}$
from the normalization condition of pion
kinetic term. As for the nucleon mass, we prefer to using a
theoretical soliton mass instead of the physical mass, since it
respects the energy-momentum sum rule at the initial energy scale.
To compare our predictions with the high energy data, we must
take account of the $Q^2$-evolution of distribution functions.
The initial energy scale necessary here is simply taken to be
$Q_{init}^2 = {(0.5 \,\mbox{GeV})}^2$, which is close to the
Pauli-Villars mass square. For solving the DGLAP evolution
equation at the NLO,
we use the Fortran Program provided by Saga group [10].

\begin{figure}[h]
\caption{The longitudinally polarized distribution in comparison
with the transversity distribution for $u$- and $d$-quarks.
The theoretical distributions at $Q^2 = Q_{init}^2$ are
evolved to $Q^2 = Q_{evol}^2$ by solving the NLO evolution equations.}
\vspace{6mm}
\begin{minipage}[t]{7.5cm}
\epsfysize=68mm
\epsfbox{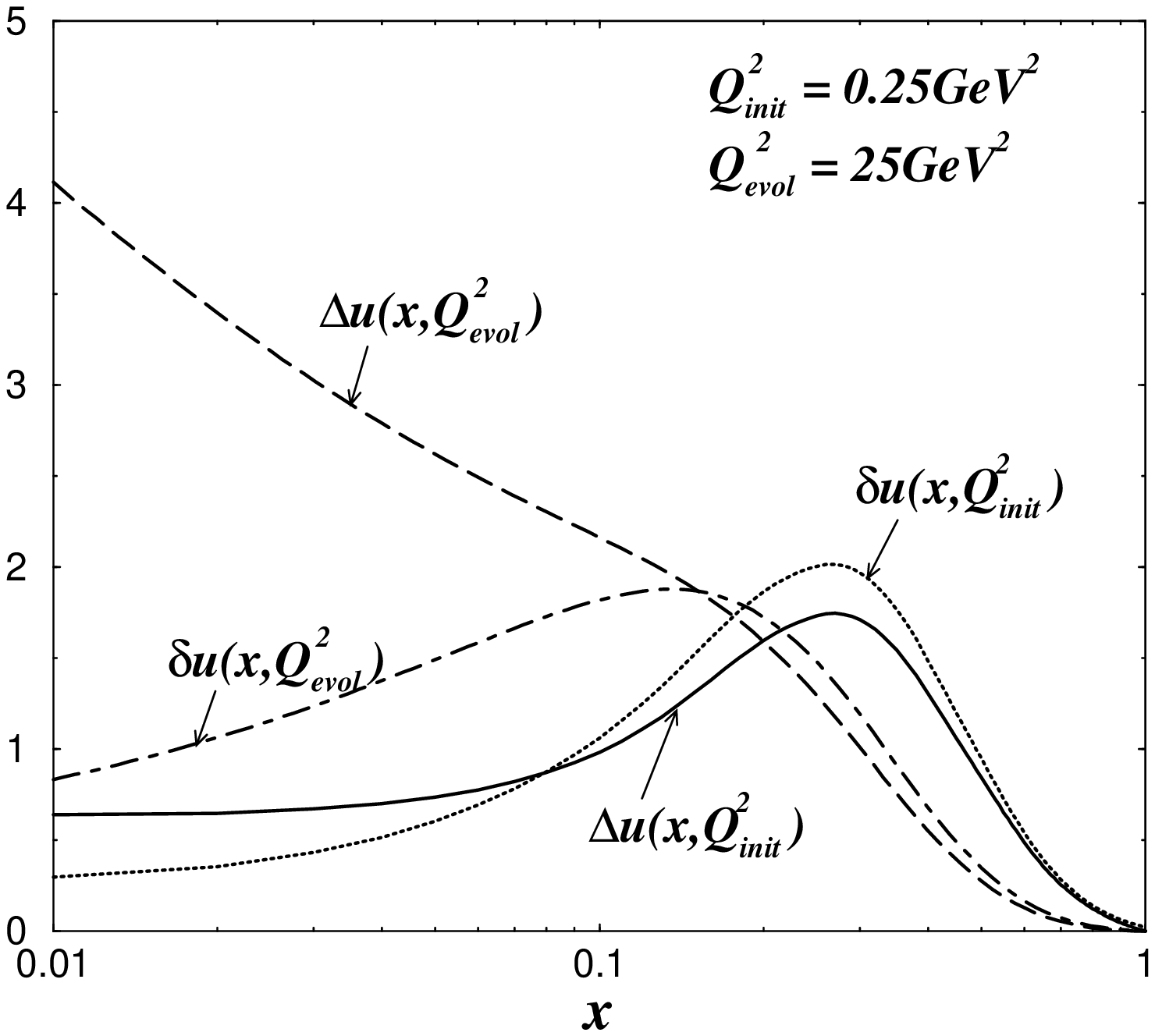}
\end{minipage}
\begin{minipage}{1cm}
\end{minipage}
\begin{minipage}[t]{7.5cm}
\epsfysize=68mm
\epsfbox{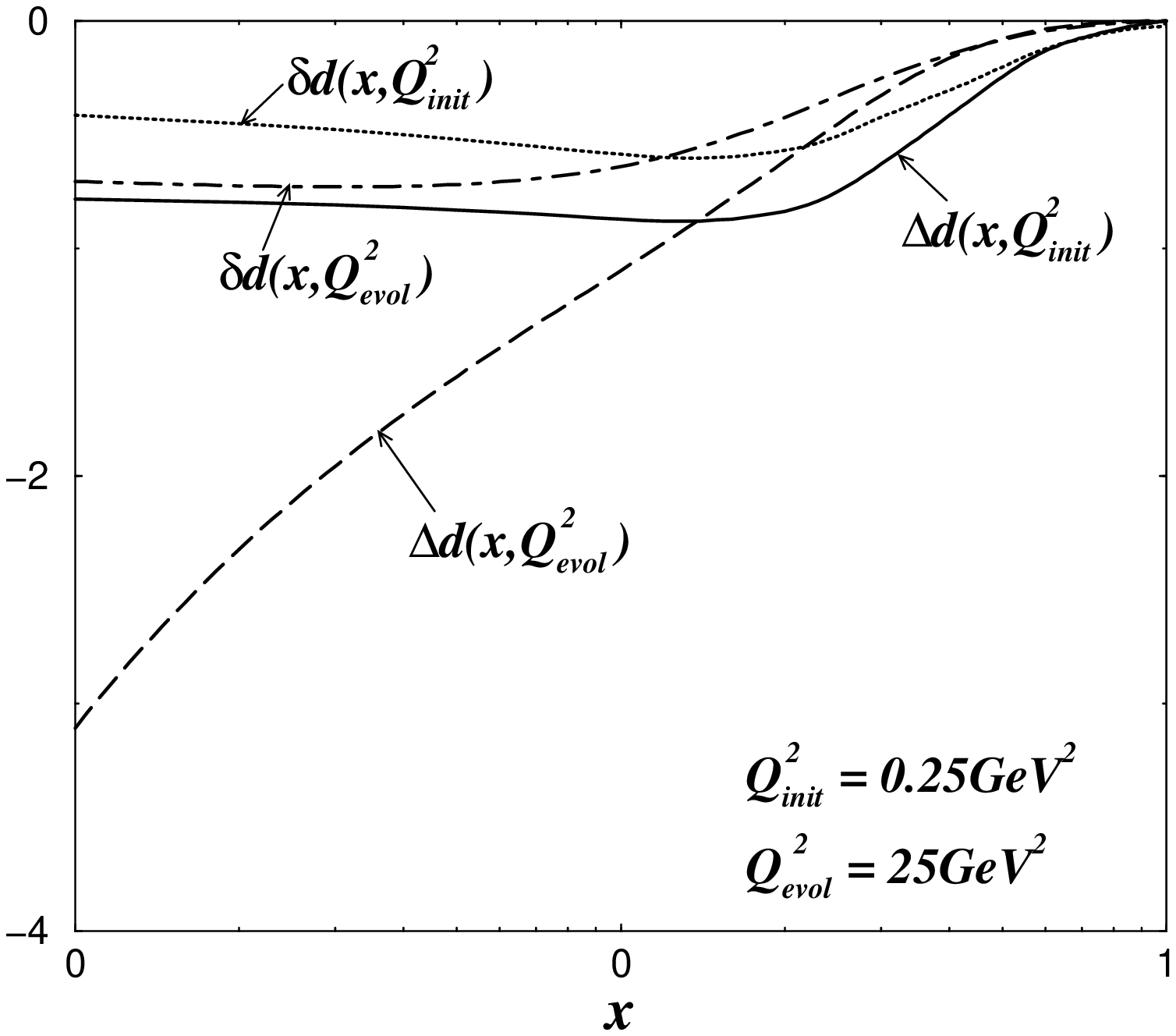}
\end{minipage}
\end{figure}

\vspace{-8mm}
\begin{figure}[h]
\caption{The longitudinally polarized distribution in comparison
with the transversity distribution for $\bar{u}$- and $\bar{d}$-quarks.
The theoretical distributions at $Q^2 = Q_{init}^2$ are
evolved to $Q^2 = Q_{evol}^2$ by solving the NLO evolution equations.}
\vspace{6mm}
\begin{minipage}[t]{7.5cm}
\epsfysize=68mm
\epsfbox{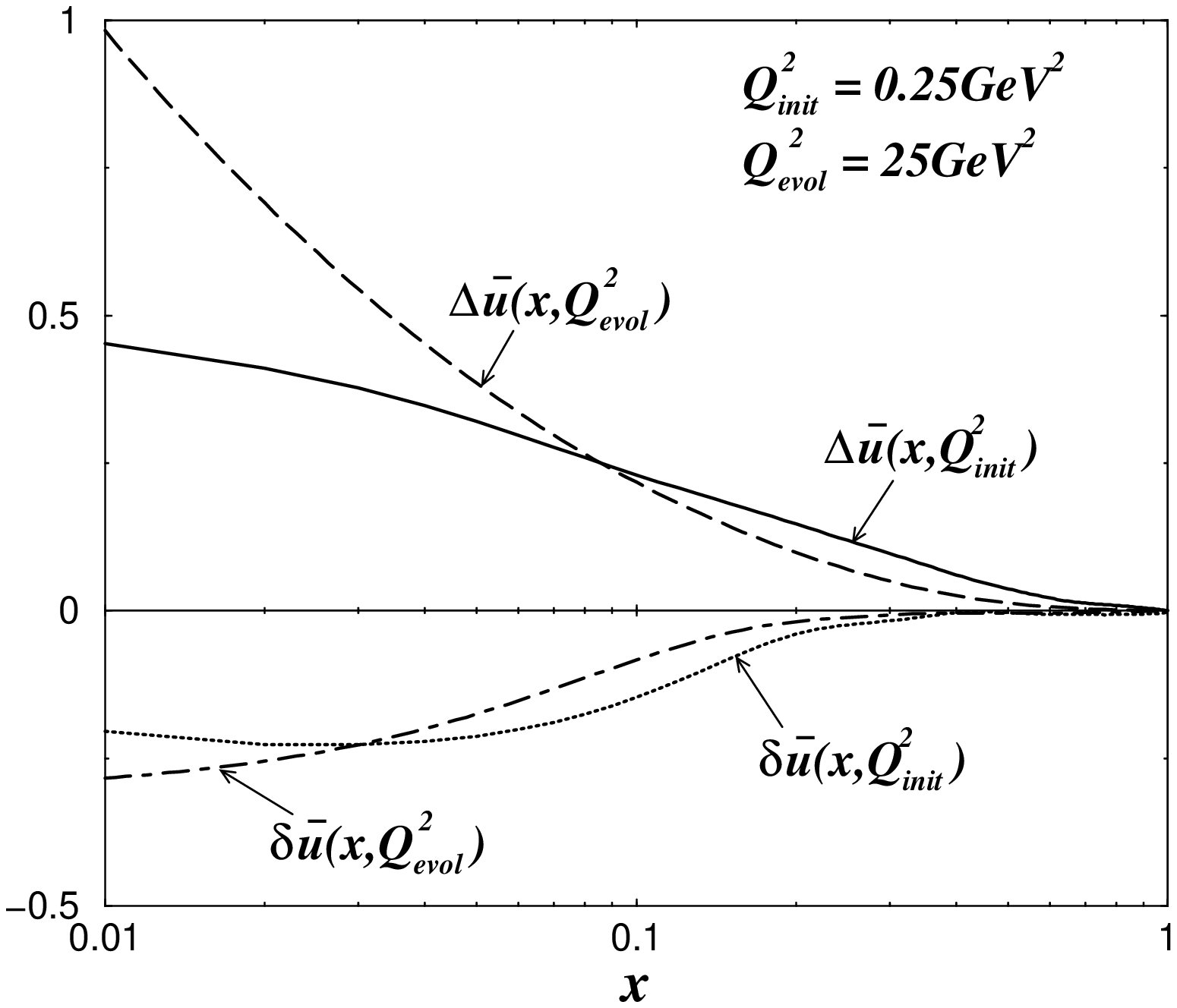}
\end{minipage}
\begin{minipage}{1cm}
\end{minipage}
\begin{minipage}[t]{7.5cm}
\epsfysize=68mm
\epsfbox{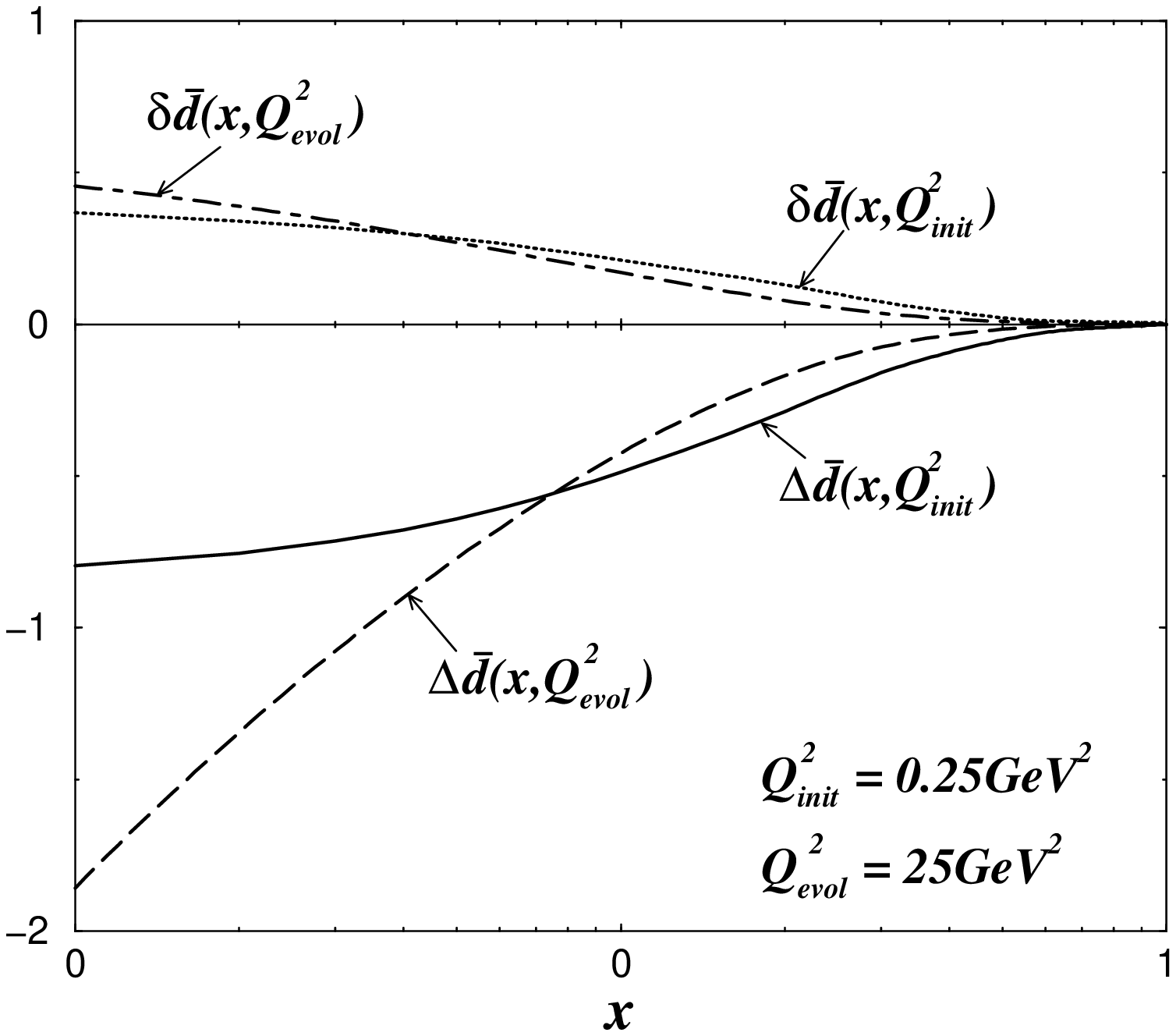}
\end{minipage}
\end{figure}

Now we show in Fig.1 and Fig.2 some of our numerical results for
quark distribution functions. Here $\Delta u (x)$ and $\delta u (x)$
respectively stand for the longitudinal and transversity
distributions for $u$-quark, whereas $\Delta d(x)$ and $\delta d(x)$
are the corresponding quantities for $d$-quark.
In our model, the difference between the two distributions are
sizable even at the initial (low) energy scale. A comparison with
yet-to-be-obtained high energy data must be done with care,
since the way of evolution of these two distributions are totally
different and the difference between these two distributions becomes
larger and larger as $Q^2$ increases [11,12]. A general trend is a
rapid growth of small $x$ component for the longitudinally
polarized distribution due to the coupling with gluons.
A similar tendency is also observed for the corresponding
$d$-quark distributions.
We can also give some predictions for the polarized antiquark
distribution functions. They are shown in Fig.2.
As one can see, even the signs are
different for the longitudinal and transversity distributions
for both of $u$ and $d$ quarks.

Now we compare our predictions for the longitudinally polarized
distributions with the existing high energy data. Shown in the
upper part of Fig.3 are the prediction of the CQSM and that of
the MIT bag model for the
spin dependent structure function $g_1^p (x)$ for the proton, in
comparison with the SLAC data [13]. As usual, the theoretical
structure functions are obtained by
convoluting the QCD coefficient functions with the evolved quark
distribution functions. One sees that a general trend of the data
is well reproduced by the CQSM. The lower part of Fig.3 shows the same
quantity for the neutron. We emphasized that the prediction of the
MIT bag model for this quantity is negligibly small even after
$Q^2$-evolution. On the other hand, the CQSM predicts sizable
magnitude of $g_1^n$ with negative sign. Although the quality of
the experimental data are not very good, they seem to support the
prediction of the CQSM.

\begin{figure}[h]
\caption{The longitudinally polarized structure functions
$g_1^p (x)$ and $g_1^n (x)$ for the proton and neutron.
The predictions of the CQSM and of the MIT bag model are
compared with the SLAC data.}
\vspace{6mm}
\begin{minipage}[t]{7.5cm}
\epsfysize=65mm
\epsfbox{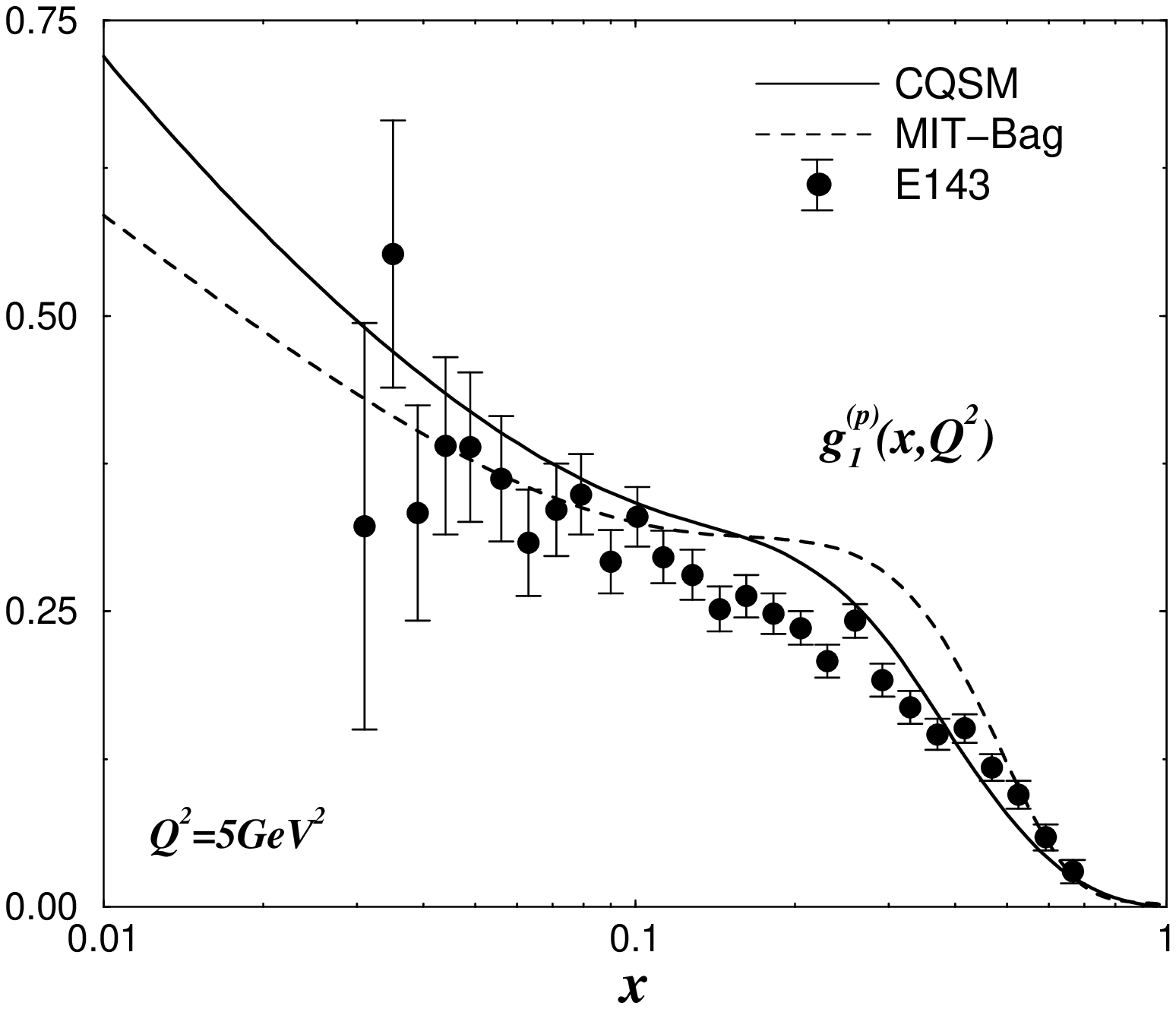}
\end{minipage}
\begin{minipage}[t]{7.5cm}
\epsfysize=65mm
\epsfbox{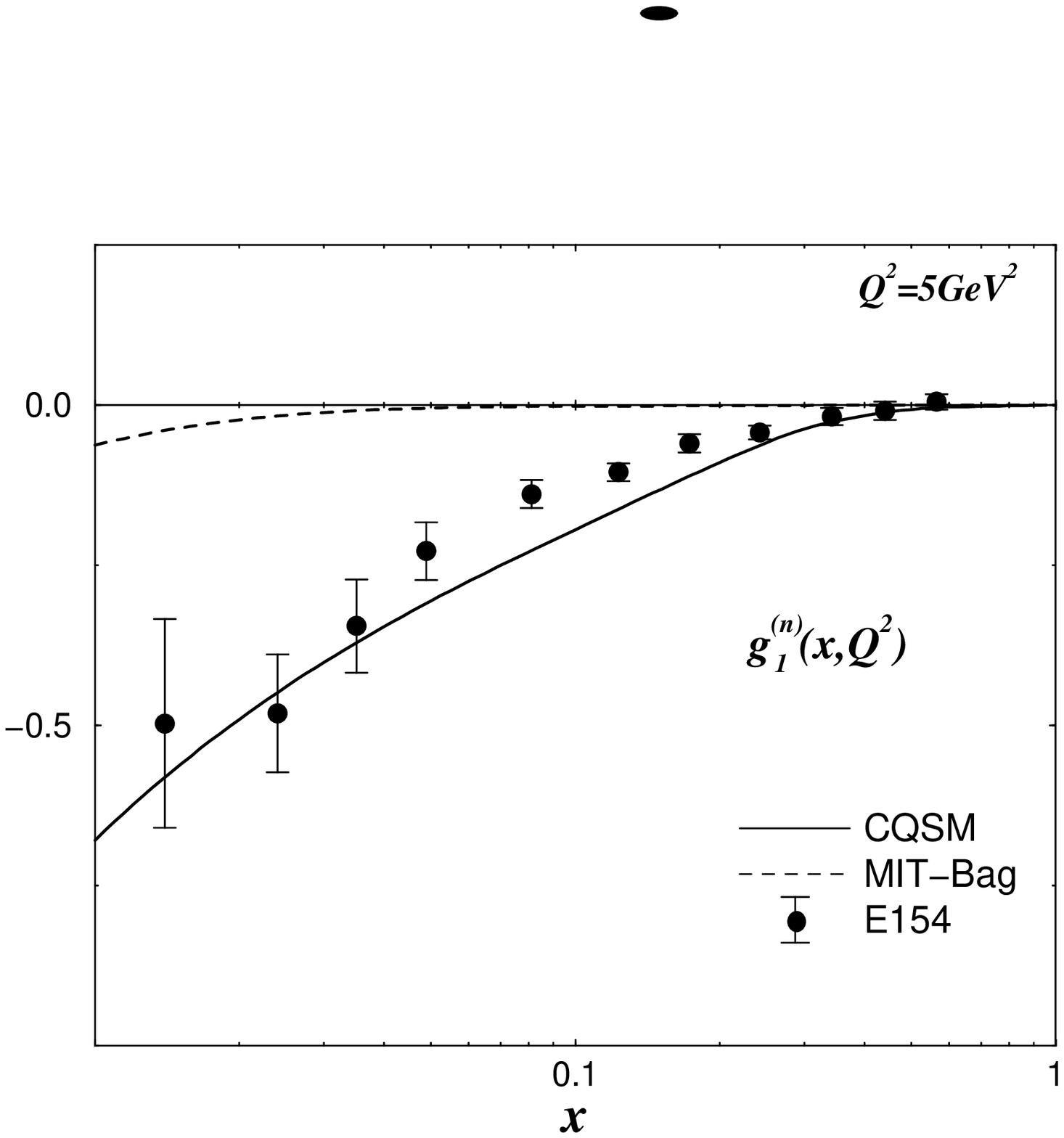}
\end{minipage}
\end{figure}

The 1st moments of the distribution functions are the simplest but
most important quantities characterizing the distributions. The 1st
moments of the longitudinally polarized distribution and the
transversity distribution are respectively called the axial and
tensor charges :
\begin{eqnarray}
 g_A^{(0,3)} &=& \int_0^1 \,d x \,\,
 \{\, [\, \Delta u (x) + \Delta \bar{u} (x) \,]
 \pm [\, \Delta d (x) + \Delta \bar{d} (x) \,] \,\} \,\, ,\\
 g_T^{(0,3)} &=& \int_0^1 \,d x \,\,\{\, 
 [\,\,\delta u (x) \,- \,\delta \bar{u} (x) \,] 
 \,\pm \,[\,\,\delta d (x) \,- \,\delta \bar{d} (x) \,] \,\} \,\, .
\end{eqnarray}
Within the framework of the nonrelativistic theory,
there is no distinction between these two charges :
\begin{eqnarray}
 g_A^{(3)} &=& g_T^{(3)} \ = \ \frac{5}{3} \,\, ,\\
 g_A^{(0)} &=& g_T^{(0)} \ = \ 1 \,\, .
\end{eqnarray}
Introduction of relativistic kinematics makes some difference,
however. In fact, the splitting of the axial and tensor charges are due
to the different sign of the lower component contribution :
\begin{eqnarray}
 g_A^{(3)} &=& \frac{5}{3} \cdot \int \,\,
 (\,f^2 \mbox{\boldmath $-$} \frac{1}{3} \,g^2 \,) \,r^2 \,dr , \ \ \ \ \ \ \  
 g_T^{(3)} = \frac{5}{3} \cdot \int \,\,
 (\,f^2 \mbox{\boldmath $+$} \frac{1}{3} \,g^2 \,) \,r^2 \,dr \,\, , \\
 g_A^{(0)} &=& \,1 \cdot \int \,\,
 (\,f^2 \mbox{\boldmath $-$} \frac{1}{3} \,g^2 \,) \,r^2 \,dr , \ \ \ \ \ \ \ 
 g_T^{(0)} = \, 1 \cdot \int \,\,
 (\,f^2 \mbox{\boldmath $+$} \frac{1}{3} \,g^2\,) \,r^2 \,dr \,\, .
\end{eqnarray}
Shown below are the numbers obtained by using typical parameters of
the bag model :
\begin{eqnarray}
g_A^{(3)} &\simeq& 1.06, \ \ \ \ \ \ \ g_T^{(3)} \simeq 1.34 \,\, ,\\
g_A^{(0)} &\simeq& 0.64, \ \ \ \ \ \ \,\,g_T^{(0)} \simeq 0.80 \,\, .
\end{eqnarray}
An important observation is that the ratios
of the isoscalar to isovector charges are $3 / 5$ for both of
the axial and tensor charges :
\begin{equation}
g_A^{(0)} / g_A^{(3)} \ = \ g_T^{(0)} / g_T^{(3)} \ = \ 3/5 \,\, .
\end{equation}
This is the case for both of the NRQM
and the MIT bag model. Now we shall argue
that this may be a limitation of valence quark models without
chiral symmetry. In fact, there is another relativistic effects,
which differentiate the axial and tensor charges. It is a dynamical
sea quark effect.

\newcommand{\lw}[1]{\smash{\lower2.ex\hbox{#1}}}
\begin{table}[h]
\caption{The axial and tensor charges of the nucleon}
\begin{center}
\renewcommand{\arraystretch}{1.0}
\begin{tabular}{|c|c|c|c|c|} \hline
 \, & \ CQSM \ & \ MIT-bag \ & $\mbox{Lattice QCD}^{*)}$ & 
 Experiment \\ \hline\hline
 \lw{$g_A^{(3)}$} & \lw{1.41} & \lw{1.06} & \lw{0.99} &
 1.254 $\pm$ 0.006 \\
 & & & & ($Q^2$-indep.) \\ \hline
 \lw{$g_A^{(0)}$} & \lw{0.35} & \lw{0.64} & \lw{0.18} &
 0.31 $\pm$ 0.07 \\
 & & & & ($Q^2$ = 10 \,$\mbox{GeV}^2$) \\ \hline\hline
 $g_T^{(3)}$ & 1.22 & 1.34 & 1.07 & -- \\ \hline
 $g_T^{(0)}$ & 0.56 & 0.80 & 0.56 & -- \\ \hline\hline
 $\ \ g_A^{(0)} / g_A^{(3)} \ \ $ & 0.25 & 0.60 & 0.18 & 0.24
 \\ \hline
 $\ \ g_T^{(0)} / g_T^{(3)} \ \ $ & 0.46 & 0.60 & 0.52 & -- \\ \hline
\end{tabular}
\end{center}
\end{table}

To see the importance of this dynamical sea quark effect, we
compare the predictions of the CQSM with those of the MIT bag model
in Table.1. A noteworthy observation is that the ratio of
the isoscalar to isovector charges are much smaller for the axial
one than for the tensor one in sharp contrast to the predictions of
the MIT bag model. It is interesting to see that this tendency is
also reproduced by a lattice calculation by Kuramashi [14].
In our opinion, this indicates an importance of chiral symmetry
as a generator of dynamical sea quark effects, and the predicted feature
is expected to be confirmed by future measurement of tensor charges.

The axial and tensor charges are generally scale dependent.
Fig.4 figure shows the next-to-leading order $Q^2$-evolution of
axial and tensor charges. As is widely known, the isovector (or
flavor-nonsinglet) axial charge is scale independent due to the
current conservation, while the flavor-singlet axial charge
has a very weak $Q^2$ dependence originating from the coupling
with gluons. Aside from about $12 \%$ overestimate of isovector
axial charge, the theoretical predictions for the axial charges
are qualitatively consistent with the experimental data.
Contrary to the axial charge, the tensor charges are known to
have strong $Q^2$ dependence [11,12]. Unfortunately, we must wait for
future experiments for obtaining any information of these
interesting observables.
Fig.5 shows the NLO evolution of the quark and gluon
polarization [15]. Here, we have assumed zero gluon polarization at
the initial energy scale. One sees that the gluon polarization
grows rapidly as $Q^2$ increases. This means that a polarized
quark is preferred to radiate a gluon with helicity parallel
to the quark spin polarization. We hope that
future experiments will give some direct information on the size
of the gluon polarization in the nucleon.

\begin{figure}[h]
\begin{minipage}[t]{7.5cm}
\caption{The NLO $Q^2$ evolution of axial and tensor charges.}
\vspace*{6mm}
\epsfysize=65mm
\epsfbox{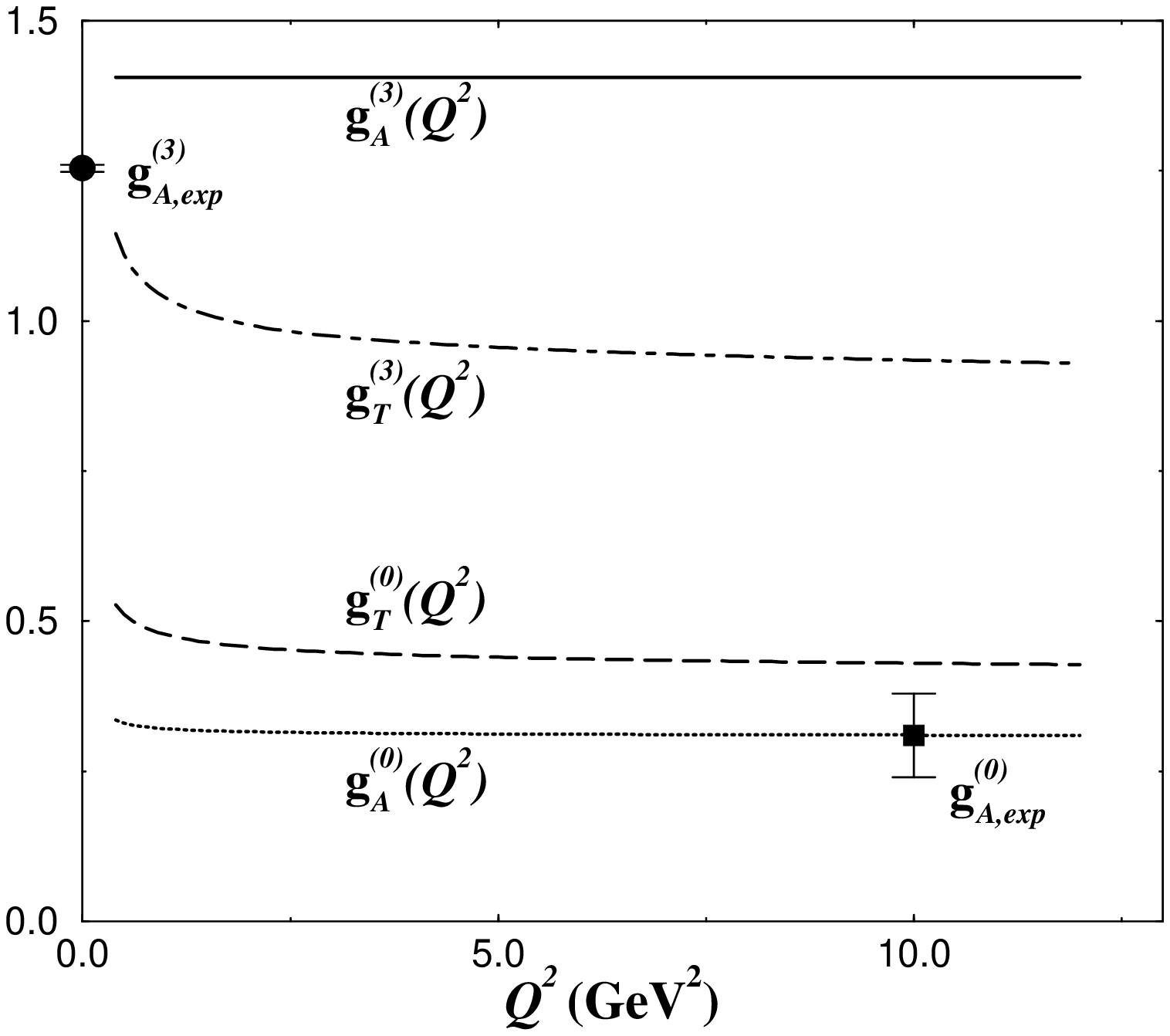}
\end{minipage}
\begin{minipage}[t]{1cm}
\end{minipage}
\begin{minipage}[t]{7.5cm}
\caption{The NLO $Q^2$ evolution of quark and gluon polarization.}
\vspace*{6mm}
\epsfysize=65mm
\epsfbox{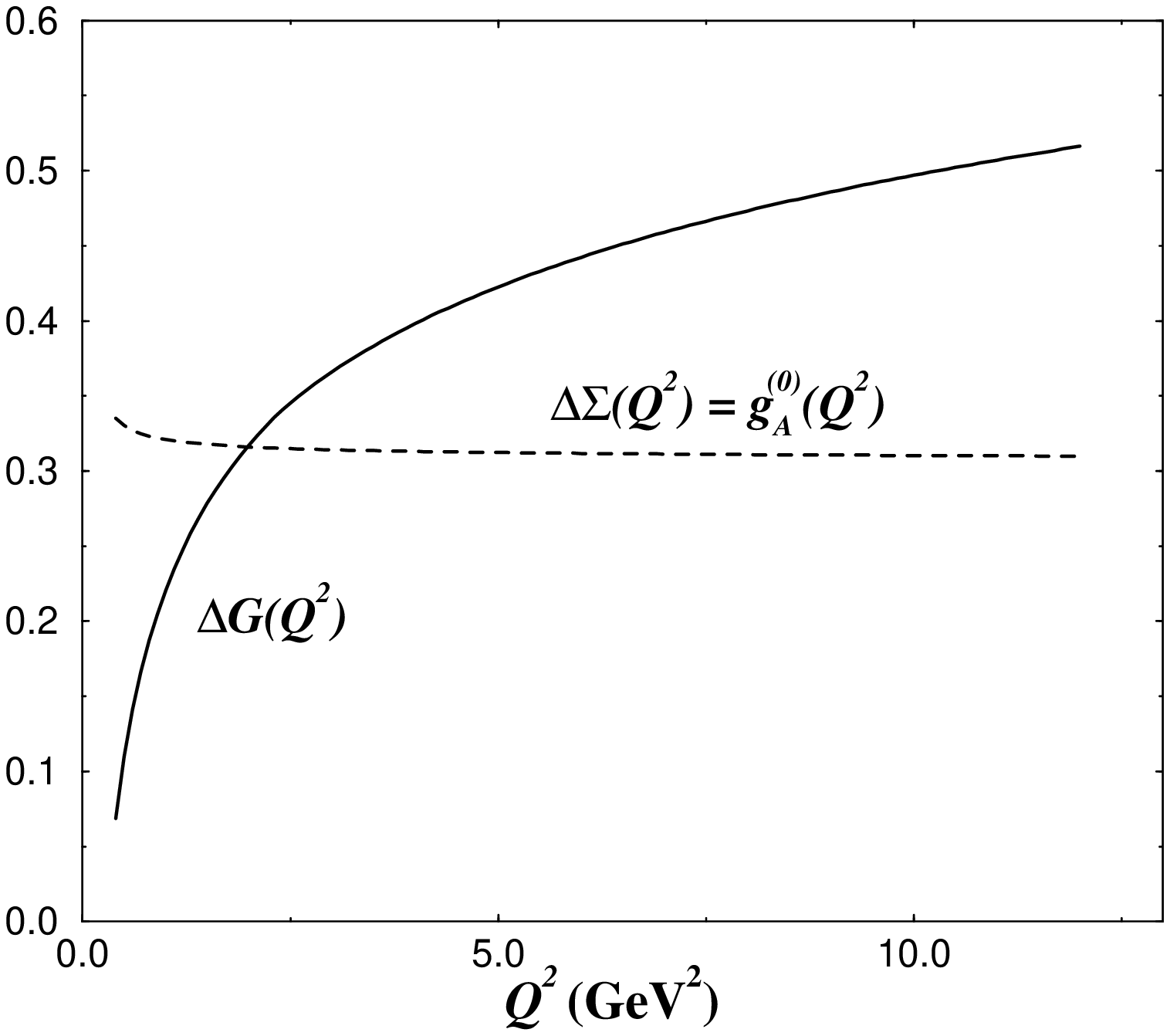}
\end{minipage}
\end{figure}

\vspace{3mm}
\noindent
\begin{Large}
{\bf 4. Summary}
\end{Large}
\vspace{3mm}

\ \ \ In summary, the CQSM naturally explains qualitative behavior
of the experimentally measured longitudinally polarized structure
functions of the proton and the neutron. Although I could not
mention in the present talk, the model also
reproduces an excess of $\bar{d}$ sea over the $\bar{d}$ sea
in the proton very naturally [6]. It also predicts qualitative
difference between the transversity distribution functions and
longitudinally polarized distribution functions.
For example, in simple valence quark models like the NRQM or the
MIT bag model, the ratios of the isoscalar to isovector charges
are just the same for both of the axial charges and the tensor
charges. On the contrary, in the CQSM, this ratio turns out to be
much smaller for the axial charges than for the tensor charges.
We have pointed out that this characteristic is also shared with
the result of lattice gauge theory. This observation then
indicates that nonperturbative QCD dynamics due to the $\chi$SB
would manifest itself in the isospin (or flavor) dependence of
high energy spin observables.

\vspace{0mm}
\section*{Acknowledgement}

\ \ \ The talk is based on the collaboration with T.~Kubota.
More detailed description on the content of the present paper
can be found in hep-ph/9809443.

%
\vspace{1mm}
\renewcommand{\baselinestretch}{0.7}
\section*{References}
\newcounter{refnum}
\begin{list}%
{[\arabic{refnum}]}{\usecounter{refnum}}
\item EMC Collaboration, J.~Aschman et al., Phys. Lett. 
{\bf B206}, 364 (1988) ; \\
Nucl. Phys. {\bf B328}, 1 (1989).
\item D.I.~Diakonov, V.Yu.~Petrov and P.V.~Pobylista, 
Nucl. Phys. {\bf B306}, 809 (1988).
\item M.~Wakamatsu and H.~Yoshiki, Nucl. Phys.
{\bf A524}, 561 (1991).
\item D.I.~Diakonov, V.Yu.~Petrov and P.V.~Pobylista,
M.V.~Polyakov and C.~Weiss,
Nucl. Phys. {\bf B480}, 341 (1996) ; Phys. Rev. {\bf D56}, 4069 (1997).
\item H.~Weigel, L.~Gamberg and H.~Reinhardt, Mod. Phys. Lett.
{\bf A11}, 3021 (1996) ;\\
Phys. Lett. {\bf B399}, 287 (1997).
\item M.~Wakamatsu and T.~Kubota, Phys. Rev. {\bf D57}, 5755 (1998).
\item J.C.~Collins and D.J.~Super, Nucl. Phys. {B194}, 445 (1982).
\item R.L.~Jaffe and Xiangdong Ji, Nucl. Phys. {bf B375}, 527 (1992).
\item M.~Wakamatsu and T.~Watabe, Phys. Lett. {\bf B312}, 184 (1993).
\item M.~Hirai, S.~Kumano and M.~Miyama, Comput. Phys. Commun.
{\bf 108}, 38 (1998) ;\\
Comput. Phys. Commu. {\bf 111}, 150 (1998).
\item X.~Artru and M.~Mekhfi, Z. Phys. {\bf C45}, 669 (1990) ;\\
Nucl. Phys. {\bf A532}, 351c (1991).
\item A.~Hayashigaki, Y.~Kanazawa and Y.~Koike, Phys. Rev. 
{\bf D56}, 7350 (1997) ;\\
S.~Kumano and M.~Miyama, Phys. Rev. {\bf D56}, 7350 (1997) ;\\
W.~Vogelsang, Phys. Rev. {\bf D57}, 1886 (1998).
\item M.~Wakamatsu and T.~Kubota, Phys. Rev. {\bf D57}, 5755 (1998).
\item K.~Abe et al., Phys. Rev. Lett. {\bf 79}, 26 (1997) ;\\
K.~Abe et al., SLAC-PUB-7753, hep-ph/9802357.
\item Y.~Kuramashi, Nucl. Phys. {\bf A629}, 235c (1998).
\item M.~Gl\"uck, E.~Reya, M.~Stratmann and W.~Vogelsang,
Phys. Rev. {\bf D53}, 4775 (1996).
\end{list}
\end{document}